\documentclass[nofootinbib, pra, twocolumn]{revtex4}
\usepackage{amsmath}
\usepackage{amssymb}
\usepackage{graphicx}
\usepackage{bm}
\usepackage{braket}
\usepackage[version=4]{mhchem}
\usepackage{color, soul}
\usepackage[dvipsnames]{xcolor}
\usepackage{bbm}
\graphicspath{ {./} }
\usepackage{float}
\usepackage{soul}
\usepackage{capt-of}
\usepackage{hyperref}

\newcommand{\beq}{\begin{equation}}
\newcommand{\eeq}{\end{equation}}

\definecolor{JM}{RGB}{4,116,149}

\definecolor{SJ}{RGB}{220,20,60}

\begin{document}

\title{Quantum Algorithm for Simulating Single-Molecule Electron Transport}

\author{Soran Jahangiri}
\affiliation{Xanadu, Toronto, ON, M5G 2C8, Canada}
\author{Juan Miguel Arrazola}
\affiliation{Xanadu, Toronto, ON, M5G 2C8, Canada}
\author{Alain Delgado}
\affiliation{Xanadu, Toronto, ON, M5G 2C8, Canada}

\begin{abstract}
An accurate description of electron transport at a molecular level requires a precise treatment of quantum effects. These effects play a crucial role in determining the electron transport properties of single molecules, such as current-voltage curves, which can be challenging to simulate classically. Here we introduce a quantum algorithm to efficiently calculate the electronic current through single-molecule junctions in the weak-coupling regime. We show that a quantum computer programmed to simulate vibronic transitions between different charge states of a molecule can be used to compute sequential electron transfer rates and electric current. In the harmonic approximation, the algorithm can be implemented using Gaussian boson sampling devices, which are a near-term platform for photonic quantum computing. We apply the algorithm to simulate the current and conductance of a magnesium porphine molecule. The simulations demonstrate quantum effects that are manifested as discrete steps in the current and conductance, in agreement with experimental and theoretical data. 
\end{abstract}

\maketitle

\section{Introduction}

The transfer of electrons through molecules is an essential component of many basic processes in physics, chemistry, and biology~\cite{kuznetrsov1995charge, misra2018intramolecular}. Single-molecule junctions have been widely used as a paradigm to investigate electron transport~\cite{evers2020advances, xu2003measurement, thomas2019understanding, burzuri2016sequential, seldenthuis2008vibrational}, with the perspective of building nanoelectronic devices composed of molecular building blocks~\cite{ratner2013a}. This concept is motivated by the growing need for miniaturizing electronic devices and by the diversity in the electronic properties of molecules, which can promise new functionalities~\cite{cuevas2010molecular}. When the size of a conducting component is reduced to the atomic scale, electron transport is dominated by quantum mechanical effects, which can lead to significant deviations from Ohm's law~\cite{cuevas2010molecular}.

The strong correlations between the tunneling electrons and the molecular degrees of freedom in single-molecule junctions are manifested as anomalous features in the current-voltage curves of these systems. Suppression of current at low bias voltages and emergence of sidebands in the differential conductance are examples of anomalous behavior that has been observed in several experiments~\cite{leturcq2009franck, burzuri2014franck, thomas2019understanding}. These features cannot be explained by single-particle models based on Landauer's theory, which assume an elastic transport mechanism and consider the junction as a scattering center~\cite{thoss2018perspective}. In the limit of weak coupling between the molecule and the electrodes, the tunneling electrons interact with the molecule's vibrational degrees of freedom. This leads to an inelastic electron transport mechanism which results in sequential steps in the current due to the quantized nature of the molecule's vibrational energy levels~\cite{koch2006theory, bevan2018relating}. In this regime, where the vibrational-electronic (vibronic) coupling is strong, the theoretical calculation of the electron transfer rates is challenging for large molecules due to the exponential increase in the number of vibrational states with the number of modes and the cutoff for the maximum number of vibrational quanta~\cite{huh2015boson, jnane2020analog, quesada2019franck}. Efficient theoretical simulations that account for the quantum effects are necessary to understand the mechanism of electron transport and to enable improved development of molecular electronic devices. 

This paper introduces a quantum algorithm for simulating electron transport in weakly-coupled single-molecule junctions. Given as input the vibrational normal modes and frequencies of the electronic ground states of the neutral and charged molecule, the algorithm can efficiently estimate the molecular vibrational density of states, which can be used to compute electron transfer rates and current. We explain the theoretical foundations of electron transport in these systems and describe our algorithm by focusing on its implementation using a photonic quantum computer. We conclude by applying the algorithm to compute current-voltage and conductance curves in a magnesium porphine single-molecule junction.

\section{Theory}

\begin{figure}[]
\includegraphics[width=1 \columnwidth]{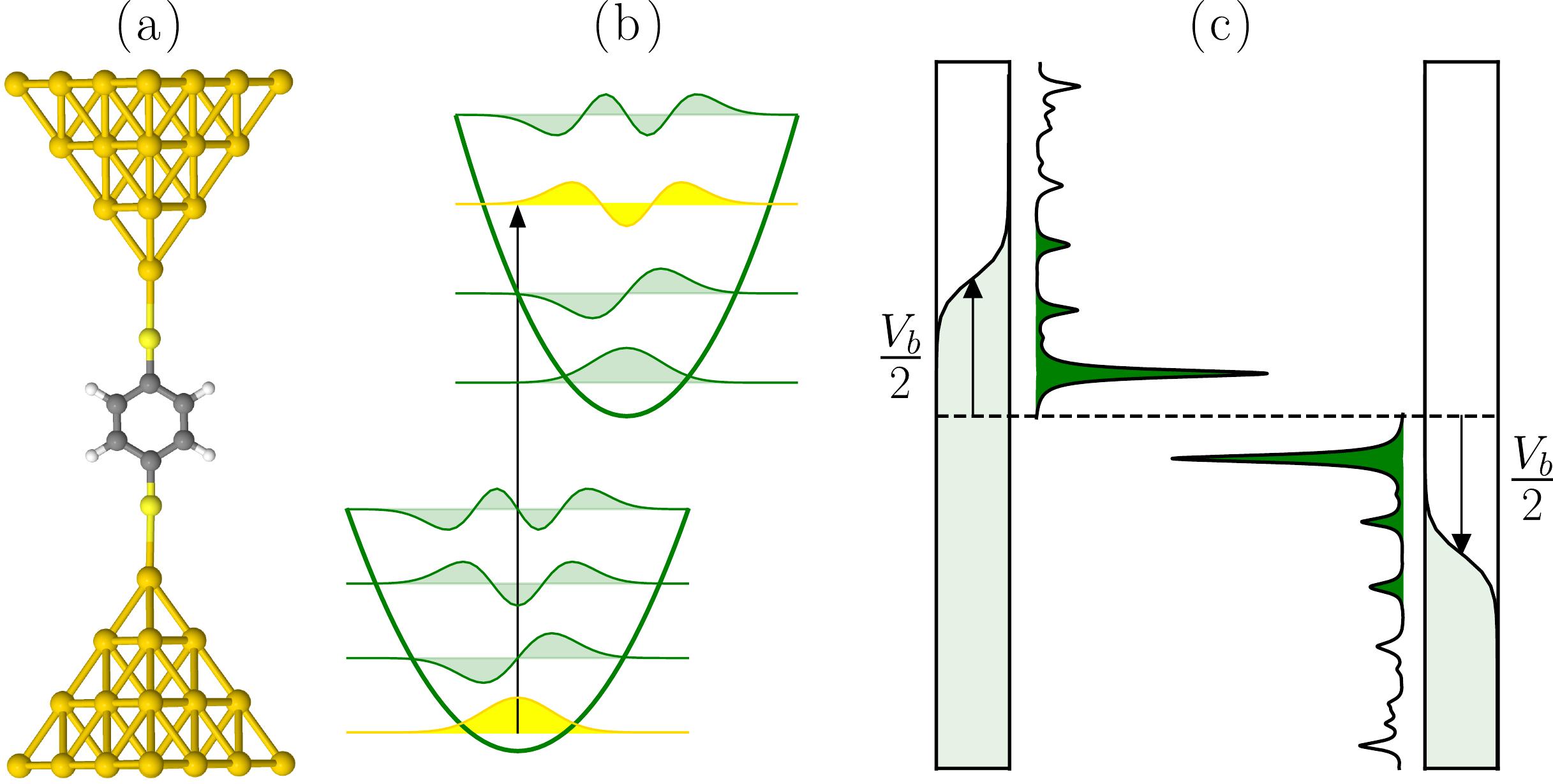}
\centering
\caption{(a) Schematic representation of a single-molecule junction bridging two electrodes. A bias voltage ($V_b$) is applied to the source and drain electrodes and a gate voltage ($V_g$) is applied to the molecule. (b) Harmonic potential energy curves for a diatomic molecule undergoing a transition between two different electronic states. The transition occurs between different vibrational states with probabilities given by the molecular density of states (DOS). (c) The bias voltage applied to the electrodes changes the Fermi level. The electrons transfer to the molecule when their energy matches the energy of the allowed transitions. These are determined by the molecular DOS illustrated by their characteristic curve in the figure. The shaded area in the molecular DOS represents the result of the integration in Eq.~\eqref{Eq: ks} to compute the electron transfer rates.}\label{fig:energy}
\end{figure}

Electron transport in single-molecule junctions is modeled with current-voltage curves that are obtained by measuring the rate of electron transport as a function of the applied bias voltage ($V_b$). The experimental setup for such measurements is composed of a molecule bridging two conducting electrodes, as illustrated in Fig.~\ref{fig:energy}. A third gate electrode is applied in some experiments to adjust the conductance of the molecule~\cite{cuevas2010molecular}.

The electron transport process in the weak coupling regime is described in terms of rate equations~\cite{mitra2004phonon, koch2005franck, koch2006theory, seldenthuis2008vibrational, bevan2018relating, thomas2019understanding}. In this formalism, the current passing through the junction is given by electron transfer rates for the charging and discharging processes~\cite{thomas2019understanding}
\beq\label{Eq: I}
I = e \frac{k_{S} k_{D}}{k_{S} + k_{D}},
\eeq
where $e$ is the electron charge, $k$ is the electron transfer rate, and $S$ and $D$ represent the source and drain electrodes, respectively. For simplicity, this expression neglects electron transfer from molecule to source electrode and from drain electrode to molecule, but these can also be considered, as explained in the Supplemental Material. 

The electron transfer rates in Eq.~\eqref{Eq: I} depend on the applied voltage, the strength of the coupling between the electrodes and the molecule, and the available energy levels of the molecule. As a function of voltage, the transfer rate from the source electrode to the molecule can be written as~\cite{thomas2019understanding, bevan2018relating}

\beq\label{Eq: ks}
k_S (V) = \frac{2\pi}{\hbar} \Gamma_S \int d\epsilon\;f(\epsilon, V) D_{r}(\epsilon) ,
\eeq
where $\Gamma_S$ is the electrode-molecule coupling, $D_{r}(\epsilon)$ represents the molecular density of states (DOS) associated with the reduction process ~\cite{thomas2019understanding, bevan2018relating}, $f(\epsilon, V)$ is the Fermi-Dirac distribution for the occupation of the electronic states in the electrode for a given voltage $V$, and $\epsilon$ is the energy.

The applied voltage aligns the energies of the electrons in the electrode with the energy of the allowed vibronic transitions between the molecular states. The transfer process occurs when the electron energies in the contact match the energies of the molecular transitions. This process is schematically shown in Fig.~\ref{fig:energy}. Similarly, the electron transfer rate from the molecule to the drain electrode is defined in terms of the molecular DOS associated with the oxidation process $D_{o}(\epsilon)$ and the corresponding coupling constant $\Gamma_D$ as~\cite{thomas2019understanding, bevan2018relating}
\beq\label{Eq: ko}
k_D (V) = \frac{2\pi}{\hbar} \Gamma_D \int d\epsilon\; [1 - f(\epsilon, V)] D_{o}(\epsilon) .
\eeq
The functions $D_{r}(\epsilon)$ and $D_{o}(\epsilon)$ depend on the populations of the vibrational states of the oxidized and reduced states and the probabilities of transitions between the vibrational energy levels of these different charge states. We write the general expression for the molecular DOS as
\beq\label{Eq: dred}
D(\epsilon) = \sum_{\bm{n}, \bm{m}} p(\bm{n}) F_{\bm{n}, \bm{m}} \delta(\epsilon - \epsilon_{\bm{n}, \bm{m}}),
\eeq
where $\bm{n}$ and $\bm{m}$ respectively represent the vibrational states of the initial and final electronic states, $p(\bm{n})$ is the probability of finding the molecule in the initial state $\bm{n}=(n_1, n_2, \ldots, n_m)$ with $n_i$ being the number of vibrational quanta in mode $i$, $F_{\bm{n}, \bm{m}}$ is the Franck-Condon factor, and $\delta$ is the Dirac delta function. We assume that the probability $p(\bm{n})$ is given by a thermal equilibrium distribution. The summations in Eq.~\eqref{Eq: dred} are over all possible vibrational states of the molecule. The number of transitions associated with these states scales combinatorially with the size of the molecule~\cite{ezSpectrum, jahangiri2020quantum}. Here, we introduce a quantum algorithm to efficiently estimate the molecular DOS and the electron transfer rates.

\section{Quantum algorithm}

We describe the algorithm in the context of photonic quantum computing, where optical modes can be conveniently used to directly represent the vibrational modes of a molecule. We further assume that the molecular vibrations can be modeled within the harmonic approximation. This simplification is made for convenience, and can be relaxed using the approaches discussed in~\cite{sawaya2019quantum} for cases where anharmonic effects are important, although these have been formulated for qubit-based quantum computers. During the vibronic transition that occurs when the molecule accepts an electron, the vibrational modes change according to the Doktorov transformation~\cite{doktorov1977dynamical}, which is represented by a corresponding Doktorov operator $\hat{U}_{\text{Dok}}$. It can be decomposed in terms of displacement $\hat{D}(\bm{\alpha})$, squeezing $\hat{S}(\bm{r})$, and linear interferometer $\hat{R}(U_L)$, $\hat{R}(U_R)$ operations as~\cite{quesada2019franck, huh2015boson}
\beq\label{Eq: Dok_Gaussian}
\hat{U}_{\text{Dok}} = \hat{D}(\bm{\alpha})\hat{R}(U_L) \hat{S}(\bm{r}) \hat{R}(U_R),
\eeq
where $U_L$ and $U_R$ are unitary matrices, $\bm{r}$ is a vector of squeezing parameters, and $\bm{\alpha}$ is a vector of displacements. The parameters of these gates are obtained from the vibrational normal modes and frequencies, and the optimized geometry of the different charge states of the molecule. More details regarding the definition of these gates and the method for obtaining the parameters is presented in the Supplemental Material. These transformations are applied to the initial state of the device and the output state can be measured in the photon-number basis to sample vibrational energies~\cite{huh2015boson}. Starting from a state $\ket{\bm{n}}=\ket{n_1, n_2, \ldots, n_m}$, where $n_i$ denotes the number of photons in mode $i$, the probability of observing a state $\ket{\bm{m}}=\ket{m_1, m_2, \ldots, m_m}$, is given by
\beq \label{Eq: GBS_Prob0}
p(\bm{m}|\bm{n}) = \left | \bra{\bm{m}}\hat{U}_{\text{Dok}}\ket{\bm{n}} \right |^2,
\eeq
which is precisely the Franck-Condon factor $F_{\bm{n}, \bm{m}}$ in Eq.~\eqref{Eq: dred}. The displacement, squeezing, and linear interferometer gates are Gaussian operations~\cite{weedbrook2012gaussian, serafini2017quantum}, meaning that the output state is also Gaussian. Since the only measurement required is photon counting, the algorithm can be implemented using Gaussian boson sampling devices, which are actively being pursued as platforms for near-term photonic quantum computing~\cite{hamilton2017gaussian, bromley2020applications, bradler2018gaussian, arrazola2018using, banchi2019molecular, jahangiri2020point, schuld2020measuring} and have recently been used in a demonstration of photonic quantum advantage~\cite{zhong2020quantum}.

Each photon pattern generated by the device corresponds to a specific transition between the vibrational energy levels of the initial and final molecular electronic states. The energy of the transition corresponding to given initial and final photon patterns can be computed as
\beq\label{Eq: energy}
E(\bm{n}, \bm{m}) = \hbar\sum_{k=1}^{N_\mathrm{vib}} \left ( m_k\omega'_k-n_k\omega_k \right ),
\eeq
where $\omega_k$ is the frequency of the $k$-th vibrational mode of the initial electronic state, $\omega'_k$ is the frequency for the final electronic state, and $N_\mathrm{vib}$ is the total number of vibrational modes.

Now consider the following approximation of the transfer rate in Eq.~\eqref{Eq: ks}.
The region of integration can be restricted to a finite interval $[\epsilon_{\min}, \epsilon_{\max}]$ and divided into $M$ intervals of width $2\Delta=(\epsilon_{\max}-\epsilon_{\min})/M$ with central energy $\epsilon_i=\epsilon_{\min}+(2i-1)\Delta$. We can then write

\begin{align}
k_S (V) &= \frac{2\pi}{\hbar} \Gamma_S\times\nonumber\\
&\sum_{i=1}^M\sum_{\bm{n}, \bm{m}} p(\bm{n}) p(\bm{m}|\bm{n})  \int_{\epsilon_{i}-\Delta}^{\epsilon_i+\Delta} d\epsilon\;f(\epsilon, V) \delta(\epsilon - \epsilon_{\bm{n}, \bm{m}}).
\end{align}
We can approximate the Fermi-Dirac function by a constant $\bar{f}(i, V)$, which we set equal to its average over the integration region
\beq
\bar{f}(i, V)=\frac{1}{2\Delta}\int_{\epsilon_{i}-\Delta}^{\epsilon_i+\Delta} d\epsilon\;f(\epsilon, V),
\eeq
and define the indicator function
\beq
\chi_i(\epsilon)= \int_{\epsilon_{i}-\Delta}^{\epsilon_i+\Delta} d\epsilon'\; \delta(\epsilon' - \epsilon).
\eeq 
We then have
\begin{align}\label{Eq: ks_final}
k_S (V)&\approx \frac{2\pi}{\hbar} \Gamma_S\sum_{i=1}^M\bar{f}(i, V)\sum_{\bm{n}, \bm{m}} p(\bm{n}) p(\bm{m}|\bm{n}) \chi_i(\epsilon_{\bm{n}, \bm{m}})\nonumber\\
&=\frac{2\pi}{\hbar} \Gamma_S\sum_{i=1}^M\bar{f}(i, V)q(i),
\end{align}
where we have implicitly defined the probability $q(i)$ of finding an energy in the $i$-th interval as
\beq\label{Eq: qi}
q(i) = \sum_{\bm{n}, \bm{m}} p(\bm{n}) p(\bm{m}|\bm{n}) \chi_i(\epsilon_{\bm{n}, \bm{m}}).
\eeq

\begin{figure}[]
\includegraphics[width=0.8 \columnwidth]{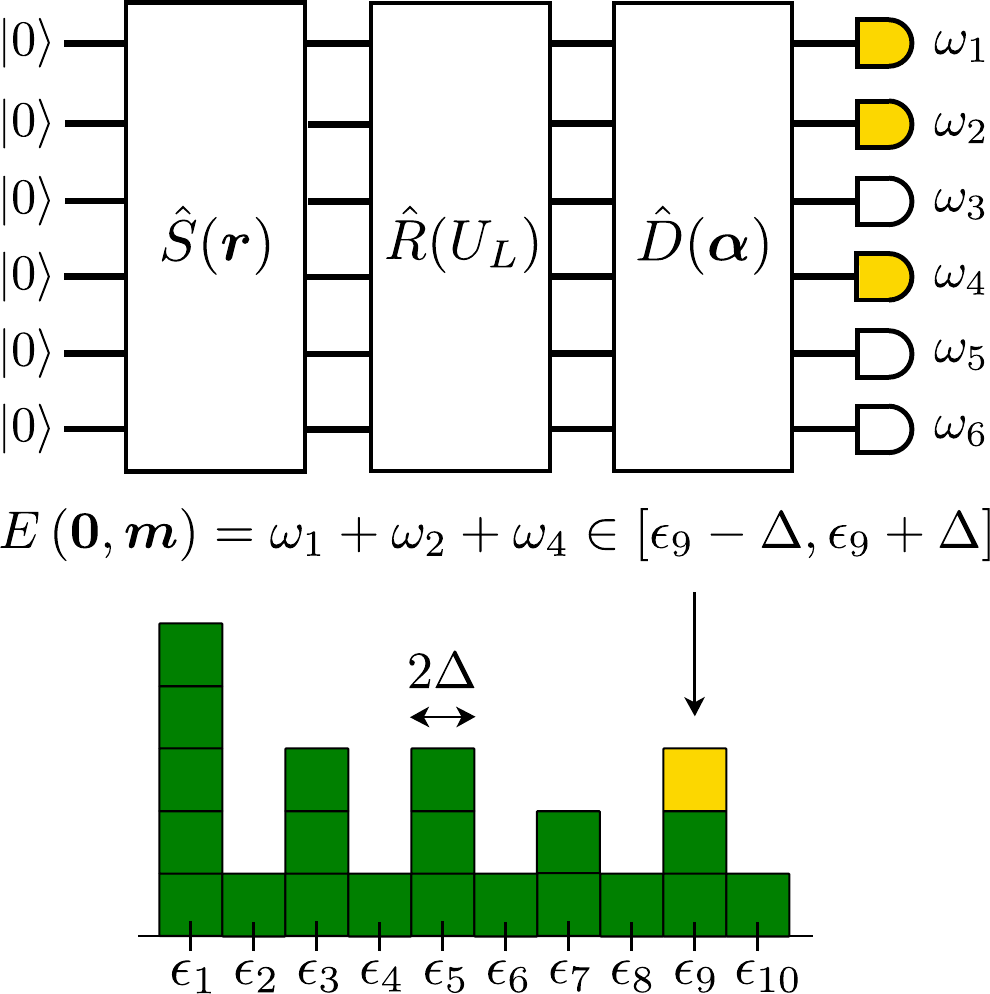}
\centering
\caption{Schematic representation of the quantum algorithm employing a photonic device with vacuum input states. The input optical modes, which correspond to the vibrational modes of a molecule, are related to the output modes according to the Doktorov transformation that is implemented using displacement $\hat{D}(\bm{\alpha})$, squeezing $\hat{S}(r)$, and linear interferometer $\hat{R}(U_L)$ operations following Eq.~\eqref{Eq: Dok_Gaussian}. The $\hat{R}(U_R)$ operation is neglected here since application of this operation to the vacuum state results in a vacuum state. Each output photon pattern generated by the device corresponds to a given energy. The samples generated by the device are used to estimate the probabilities $q(i)$ used in Eq.~\eqref{Eq: ks_final} as $q(i)=N_i/N$, where $N_i$ is the number of samples in bin $i$ and $N$ is the total number of samples.}\label{fig:gbs}
\end{figure}

By using a quantum computer to simulate a Doktorov transformation, the coarse-grained probabilities $q(i)$ can be efficiently estimated by generating samples from the output state, computing their corresponding energy, and registering the number that fall inside each interval. This is the essence of the quantum algorithm, which we describe in detail below. For simplicity, we focus on the case of zero temperature where the initial state is the vacuum in all modes. The method described in Ref.~\cite{huh2017vibronic} can be used for finite temperature simulations. The photonic quantum computer setup and the procedure for estimating the probabilities $q(i)$ are schematically shown in Fig.~\ref{fig:gbs}.\\

\begin{center}
\textbf{Algorithm}
\end{center}

\begin{enumerate}

\item Obtain the vibrational normal modes and frequencies $\omega, \omega'$ of the initial and final electronic sates of the molecule. Use these to compute the parameters $\bm{\alpha}$, $U_L$, $\bm{r}$, and $U_R$ of the Doktorov transformation $\hat{U}_{\text{Dok}}$.

\item Initialize the device to the vacuum in all modes. Apply the sequence of Gaussian gates $\hat{S}(\bm{r})$, $\hat{R}(U_L)$, and $\hat{D}(\bm{\alpha})$ that constitute the Doktorov transformation. We neglect the $\hat{R}(U_R)$ operation since the input state is vacuum for zero temperature simulations.

\item Measure the output state in the photon-number basis to obtain an outcome $\ket{\bm{m}}$. Use Eq.~\eqref{Eq: energy} to compute its corresponding energy $E(\bm{0}, \bm{m})$.

\item Repeat the above steps $N$ times.

\item Select an energy interval $[\epsilon_{\min}, \epsilon_{\max}]$ and divide it into $M$ subintervals. For each sampled energy, identify which interval it belongs to. Let $N_i$ be the number outcomes that fall in the $i$-th interval. Set $q(i)=N_i/N$. 

\item Compute the rate of the electron transfer from the source electrode to the molecule as
\beq\label{Eq:k_S_algo}
k_S (V) = \frac{2\pi}{\hbar} \Gamma_S\sum_{i=1}^M\bar{f}(i, V)q(i).
\eeq

\item Repeat the above steps to compute the rate of the electron transfer from the molecule to the drain electrode as
\beq\label{Eq:k_D_algo}
k_D (V) = \frac{2\pi}{\hbar} \Gamma_S\sum_{i=1}^M[1-\bar{f}(i, V)]q(i).
\eeq

\item Obtain the current-voltage curve by using Eq.~\eqref{Eq: I} and the rates computed in the previous step for different voltages. 

\end{enumerate}

The algorithm can also be understood as a technique to compute DOS curves. Specifically, the algorithm can produce an approximation of the DOS as a sequence of rectangular functions. 
\beq
D(\epsilon) = \frac{1}{2\Delta}\sum_{i=1}^M q(i) \chi_i(\epsilon),
\eeq
that when inserted in Eq.~\eqref{Eq: ks}, reproduces the expression of Eq.~\eqref{Eq:k_S_algo}. 
The errors in estimating the probabilities $q(i)$ follow from standard results in estimation theory. As shown explicitly in the Supplemental Material, the number of samples needed scales as $O(M/\varepsilon^2)$, where $\varepsilon$ is the error in the estimation.

We now showcase the algorithm by employing it in an example application to obtain current-voltage curves in a magnesium porphine single-molecule junction.

\section{Application}

\begin{figure*}[]
\includegraphics[width=1.8 \columnwidth]{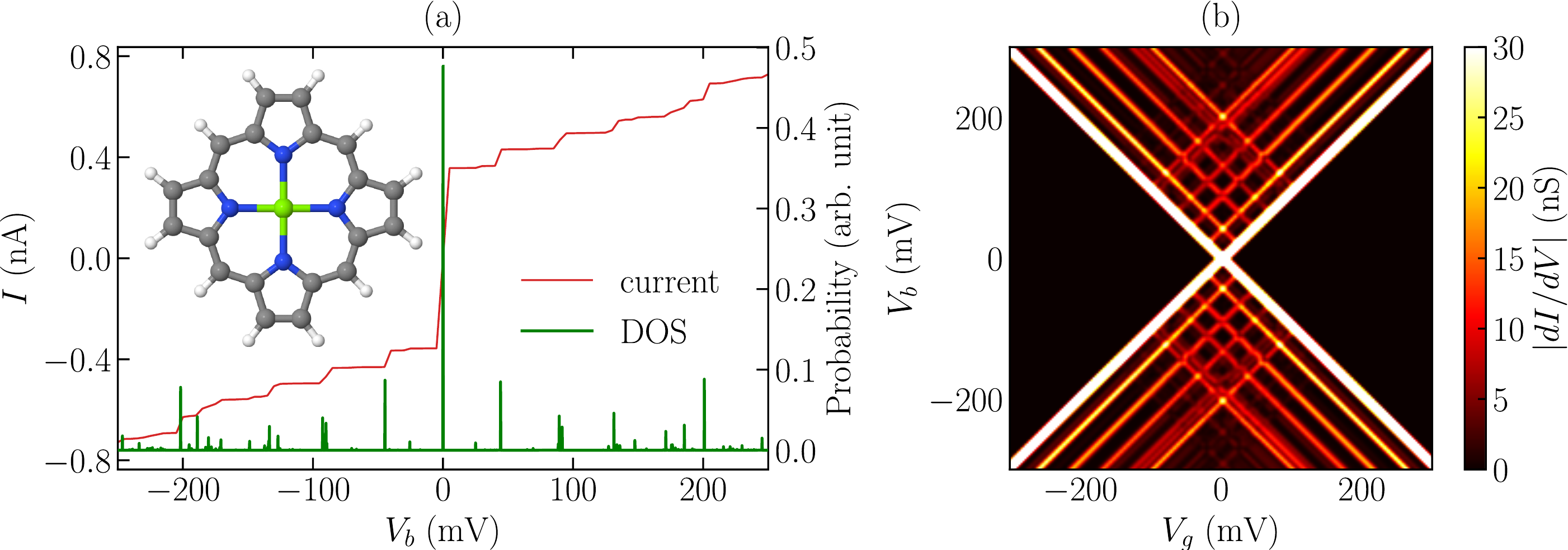}
\centering
\caption{(a) Molecular density of states (DOS) associated with the oxidation and reduction processes of magnesium porphine (MgP) computed from 5000 samples. The DOS associated with the oxidation process is plotted at the positive region of the bias voltage ($V_b$) where an electron is transferred from the molecule to the drain electrode, and the DOS associated with the reduction process is plotted at the negative bias voltage region where an electron is transferred from the source electrode to the molecule. The current-voltage curve for MgP computed from Eq.~\eqref{Eq: I} at a gate voltage ($V_g$) of $1252.6$ mV is also presented. The optimized geometry of MgP obtained from DFT calculations is shown in the inset. (b) Differential conductance of MgP computed for different bias and gate voltages. The dark color represents the suppressed current region and the bright lines represent the jumps in conductivity due to the transition between the quantized vibrational energy levels as described by the molecular DOS. The gate voltage is shifted by $-1252.6$ mV.}\label{fig:res}
\end{figure*}

The optimized geometry of the magnesium porphine (MgP) molecule is shown in Fig.~\ref{fig:res}(a). The electron transport behavior of MgP and several other porphyrins has been investigated both experimentally and theoretically~\cite{nazin2005tunneling, thomas2019understanding, perrin2011charge}. Results of the investigations indicate the effects of vibronic transitions on the electron transport of these materials as distinct bands in the current-voltage curves.

The equilibrium structures of the neutral and reduced MgP molecule and the corresponding vibrational normal modes and frequencies were obtained from density functional theory (DFT)~\cite{hohenberg1964inhomogeneous} calculations with the hybrid B3LYP~\cite{lee1988development,becke1993density,stephens1994ab} functional and the Pople basis set 6-311++G(d,p)~\cite{ditchfield1971self}. The electronic structure calculations were performed with the general atomic and molecular electronic structure system (GAMESS)~\cite{schmidt1993general,gordon2005advances}. The coupling constants $\Gamma_S$ and $\Gamma_D$ were considered to be equal and have an approximate value of $1 \times 10^{-6}$ eV~\cite{thomas2019understanding}. The MgP molecule has 105 normal modes, leading to a quantum algorithm on 105 optical modes. However, due to small amounts of squeezing and displacement, the number of photons in the output state remains small in this case, which allows us to still perform a classical simulation~\cite{quesada2020exact,quesada2020quadratic}. Samples were obtained by simulating the quantum circuits using the algorithms implemented in Strawberry Fields~\cite{killoran2019strawberry} and The Walrus~\cite{gupt2019walrus}. 

The molecular DOS associated with the oxidation and reduction of MgP, corresponding respectively to positive and negative voltages, are presented in Fig.~\ref{fig:res}(a). They are obtained from 5000 samples for each process. In both cases there is a large peak that corresponds to the transition between the ground vibrational states of the initial and final electronic states. This is followed by several smaller peaks that result from the transitions to other vibrational states.

The resulting DOS was used in each case to compute the current as a function of the bias voltage from Eq.~\eqref{Eq: I}. The resulting curve is shown in Fig.~\ref{fig:res}(a). Increasing the voltage results in a sudden change in the current due to the presence of the large distinct peak in the DOS. Further increases in the voltage lead to small jumps in the current at energies that correspond to the positions of other peaks in the DOS. The presence of such steps in the current, as a result of the vibronic transitions, has been confirmed experimentally~\cite{nazin2005tunneling, thomas2019understanding, perrin2011charge}. The differential conductance ($dI/dV$) map of MgP, computed at different bias and gate voltages, is shown in Fig.~\ref{fig:res}(b). The gate voltage $V_g$ is used to adjust the energy levels of the molecule with respect to the Fermi levels of the electrodes. The effects of the vibronic transitions are manifested as distinct bands in these diagrams, which are predicted by our algorithm.

\section{Conclusions}

Electron transport in molecular systems exhibits a rich variety of features stemming from quantum mechanical effects. A precise simulation of these phenomena is challenging since there is not a unified theory for describing the mechanism of electron transport in molecules and because the \textit{ab initio} computations that account for these effects are computationally expensive.

Here, we have introduced a quantum algorithm for computing the electric current passing through a single-molecule junction in the weak-coupling regime.
We show that a quantum computer programmed with the appropriate molecular data can be used to obtain current-voltage curves that reproduce experimentally-verified quantum effects. The algorithm is efficient: it uses circuits of polynomial depth and it produces an estimate of the transfer rates, or equivalently the molecular density of states, using a number of samples that scales polynomially with the target error in the estimation. Within the harmonic approximation, the algorithm can be implemented using Gaussian boson sampling devices, which are being actively developed as a near-term platform for photonic quantum computing. 

We showcased the properties of the algorithm by applying it to compute the current-voltage characteristics of the magnesium porphine molecule. The results demonstrate the appearance of steps in the current-voltage curve and side bands in the conductance map as a result of the quantized vibronic transitions mediating the process of electron transport. We expect the predictions to be improved by including anharmonic effects, which requires exiting the regime of Gaussian boson sampling, and by employing more accurate methods for the electronic structure calculations needed to obtain the molecular parameters. Overall, the algorithm is a first step in understanding how quantum computing can impact the development of molecular electronics. 

\section*{Acknowledgments}

We thank Nicol\'as Quesada and Maria Schuld for fruitful discussions.

\bibliographystyle{apsrev}
\bibliography{ET_references}

\end{document}


\title{Supplemental Material for:\\ 
Quantum Algorithm for Simulating Single-Molecule Electron Transport}

\author{Soran Jahangiri}
\affiliation{Xanadu, Toronto, ON, M5G 2C8, Canada}
\author{Juan Miguel Arrazola}
\affiliation{Xanadu, Toronto, ON, M5G 2C8, Canada}
\author{Alain Delgado}
\affiliation{Xanadu, Toronto, ON, M5G 2C8, Canada}

\begin{abstract}

\end{abstract}

\maketitle

\section{Electric current}

\begin{figure}[b]
\includegraphics[width=1 \columnwidth]{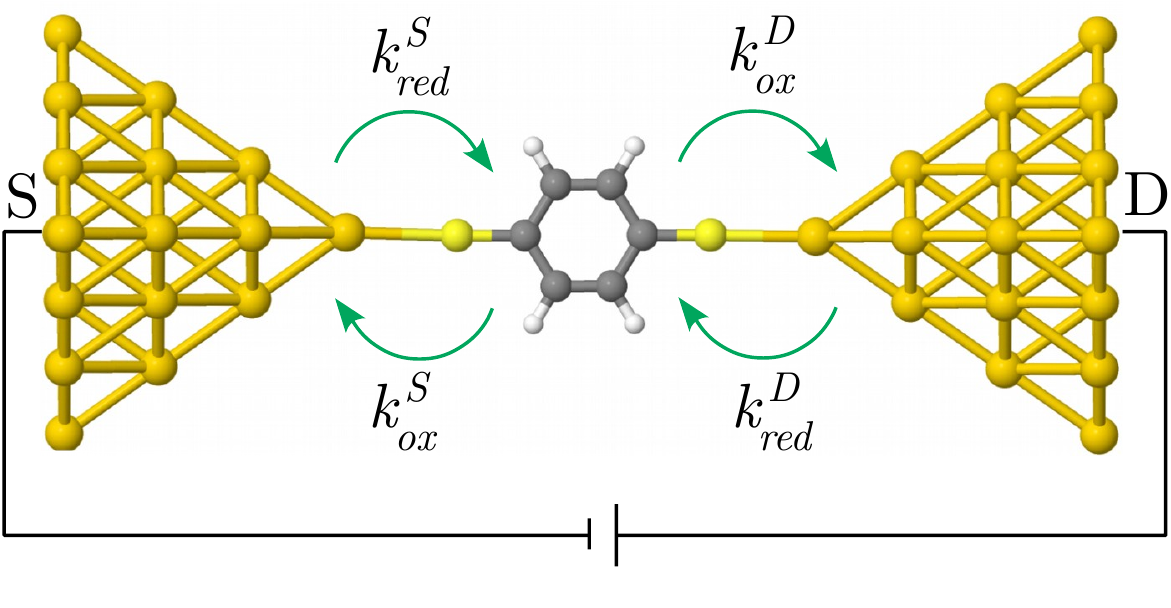}
\centering
\caption{Schematic representation of a single-molecule junction bridging a source (S) and a drain (D) electrode. The electron transfer rates in Eq.~\eqref{Eq: I2} are also defined.}\label{fig:junction}
\end{figure}

\begin{figure*}[t]
\includegraphics[width=2 \columnwidth]{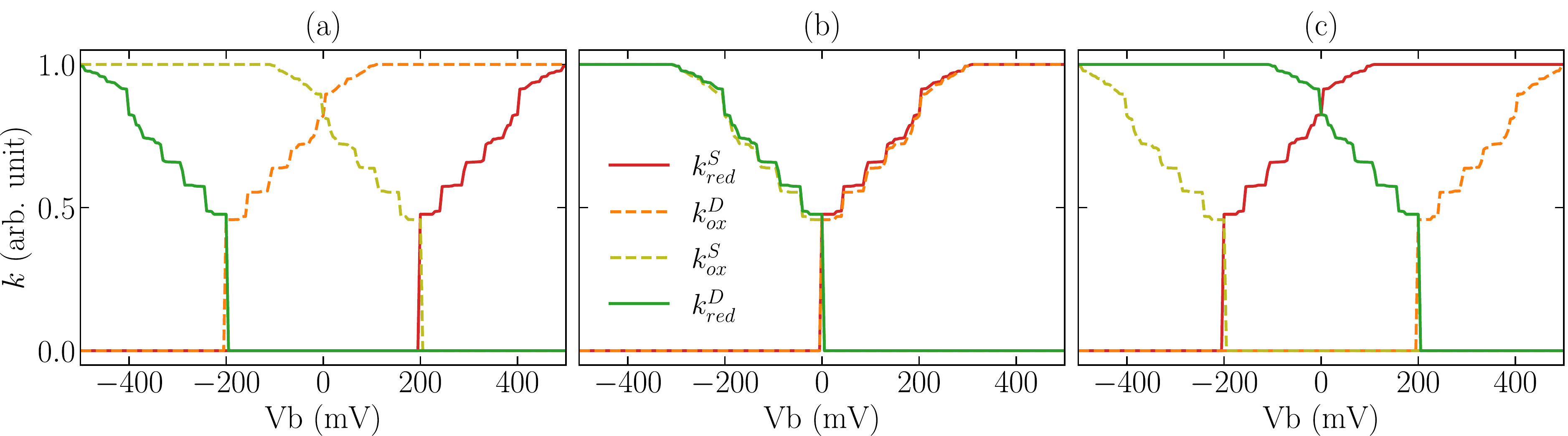}
\centering
\caption{Electron transfer rates for MgP computed as a function of the bias voltage at gate voltages of (a) -200.0 mV, (b) 0.0 mV and (c) 200.0 mV. The gate voltage is shifted by $-1252.6$ mV.}\label{fig:rate}
\end{figure*}

\begin{figure}[b]
\centering
\includegraphics[width=1.0 \columnwidth]{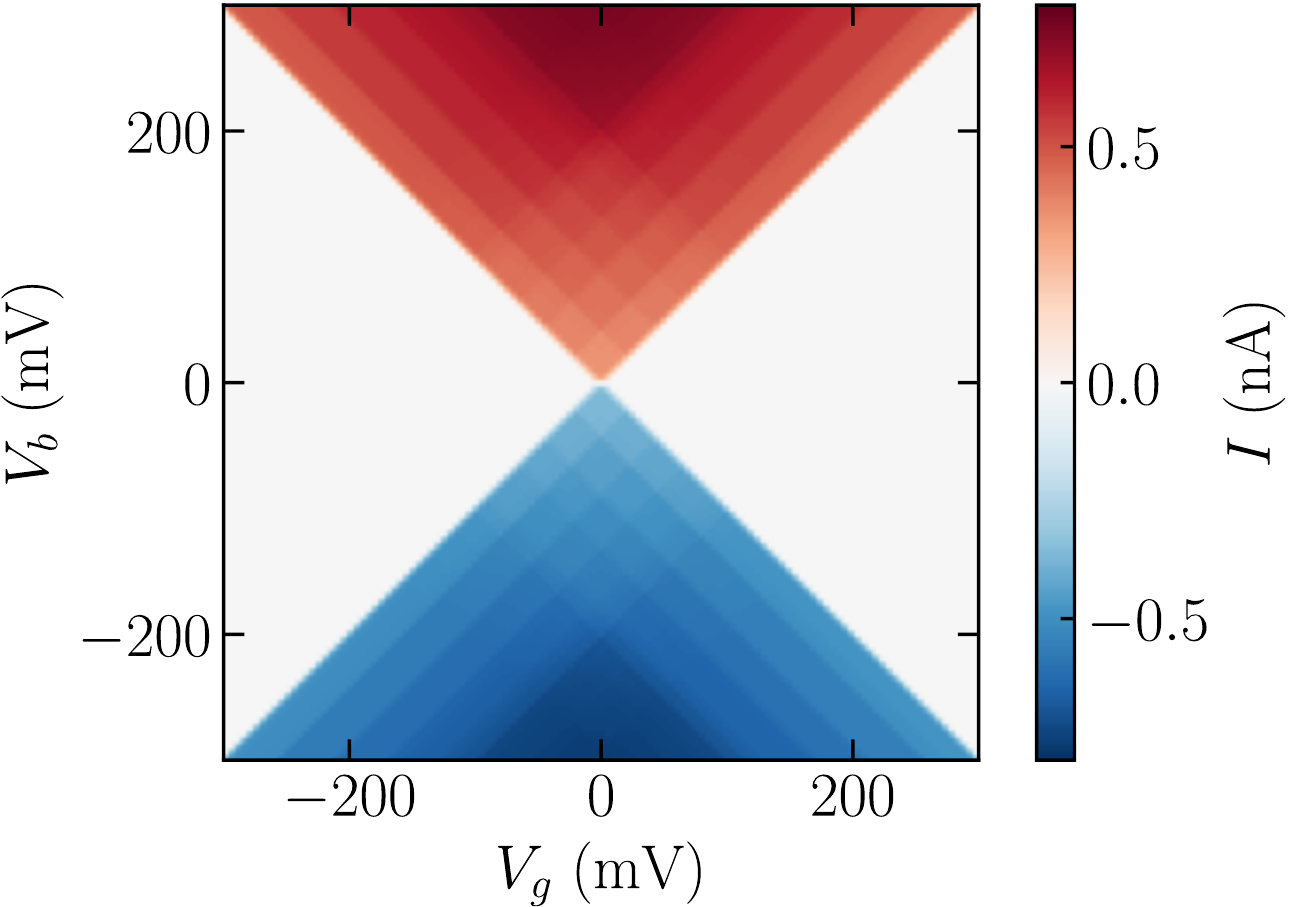}
\caption{Electric current in magnesium porphine computed as a function of bias and gate voltages. The gate voltage is shifted by $-1252.6$ mV.}\label{fig:cur}
\end{figure}

The current flowing through a molecular junction, in the limit of weak coupling between the junction and the conducting electrodes, is a result of sequential electron transfers between the molecule and the electrodes. This leads to consequent changes in the charge state of the molecule when an electron is transferred to the molecule and when the molecule loses the electron. The probability of the molecule being at these electronic states is denoted respectively by $P_{ox}$ and $P_{red}$. The current is computed from these occupation probabilities as \cite{thomas2019understanding, bevan2018relating}
\beq
I = e(P_{ox}k_{red}^S - P_{red}k_{ox}^S),
\eeq   
where $k_{a}^b$ is the electron transfer rate as defined below in Fig.~\ref{fig:junction} and $e$ is the electron charge. The occupation probabilities are time dependent and can be estimated from master equations of the form
\beq\
\frac{dP_{ox}}{dt} = -(k_{red}^S + k_{red}^D)P_{ox} + (k_{ox}^S + k_{ox}^D)P_{red}.
\eeq
The current is computed in the steady-state limit where $dP_{ox}/dt = 0$ and by recalling that $P_{ox} + P_{red} = 1$. The current is then obtained as
\beq\label{Eq: I2}
I = e \frac{k_{red}^S k_{ox}^D - k_{ox}^S k_{red}^D}{k_{red}^S + k_{ox}^D + k_{ox}^S + k_{red}^D}.
\eeq
For the magnesium porphine (MgP) molecule considered in the main text, we plot the values of the electron transfer rates as a function of the bias voltage in Fig.~\eqref{fig:rate}. The current computed from the electron transfer rates, for different bias and gate voltages, is plotted in Fig.~\eqref{fig:cur}. If we neglect oxidation at the source electrode and reduction at the drain electrode, Eq.~\eqref{Eq: I2} can be written as
\beq\label{Eq: I3}
I = e \frac{k_{S} k_{D}}{k_{S} + k_{D}}.
\eeq

\begin{figure}[b]
\centering
\includegraphics[width=0.9 \columnwidth]{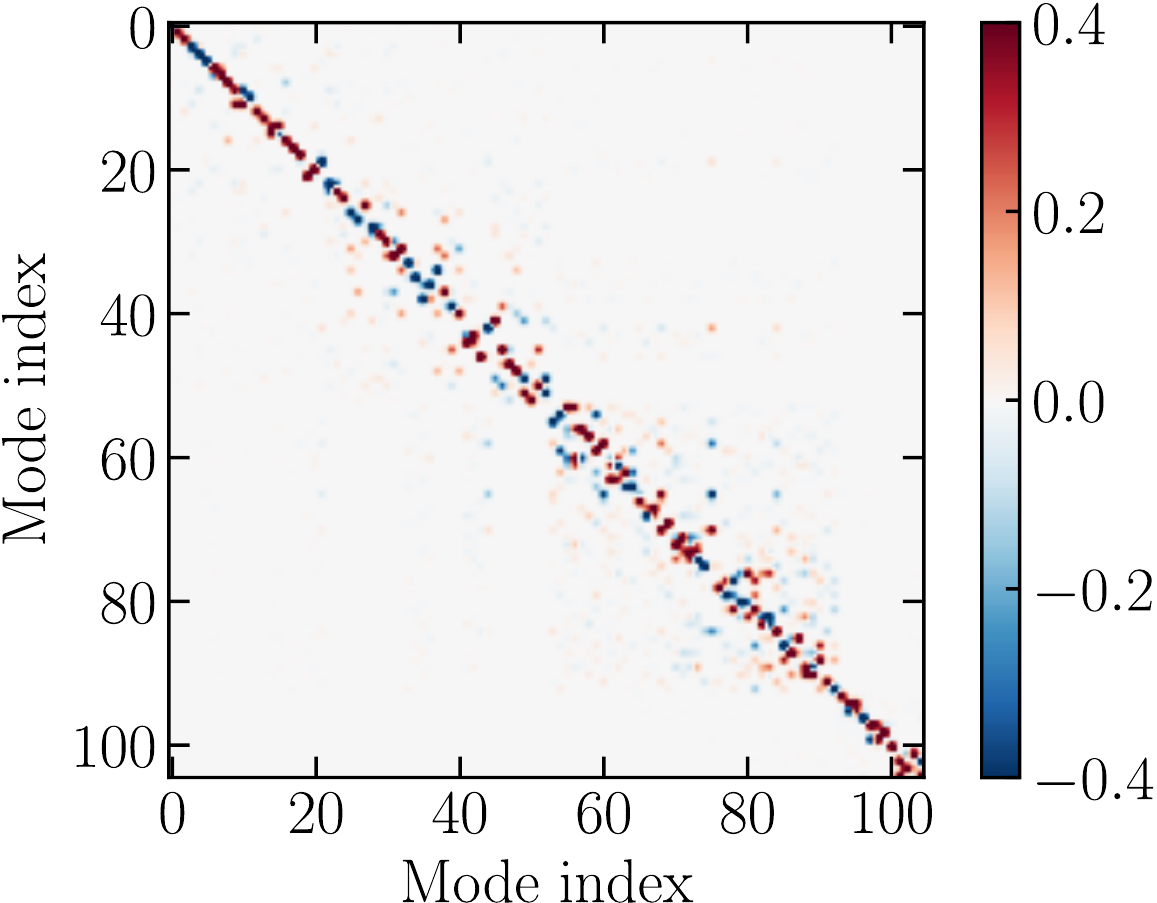}
\caption{Duschinsky matrix for magnesium porphine computed from Eq.~\eqref{Eq: UD}. The molecule has 37 atoms and 105 vibrational modes.}\label{fig:dus}
\end{figure}

\section{Photonic gates}

In photonic quantum computing, the fundamental physical systems of interest are optical modes of the quantized electromagnetic field. They are mathematically represented by a Hilbert space of infinite dimension, where a general state can be expressed as $\ket{\psi}=\sum_{n=0}^\infty c_n \ket{n}$, with $\sum_{n=0}^{\infty}|c_n|^2=1$. The basis states $\ket{n}$ are known as \emph{Fock states}, and they have the physical interpretation of a mode with $n$ photons.

The state of a multi-system can also be uniquely specified by its Wigner function. Gaussian states have a Wigner function which is a Gaussian distribution. Similarly, Gaussian gates are unitary transformations that map Gaussian states into Gaussian states. In terms of the creation and annihilation operators $a_i$ and $a_i^\dagger$ on mode $i$, a squeezing gate is given by 
\beq
\hat{S}(r_i)=\exp[r_i(a_i^{\dagger 2}-a_i^2)/2],
\eeq
a displacement gate by 
\beq
\hat{D}(\alpha_i)=\exp(\alpha_i a_i^\dagger - \alpha_{i}^* a_i),
\eeq
and the linear interferometer $\hat{R}(U)$ characterized by a unitary matrix $U$ transforms the mode operators as
\beq
\begin{pmatrix}
a_1 \\a_2\\\vdots \\a_m\end{pmatrix}\longrightarrow
\begin{pmatrix}
a_1' \\a_2'\\\vdots \\a_m'\end{pmatrix}
 = U\begin{pmatrix}
a_1 \\a_2\\\vdots \\a_m\end{pmatrix}.
\eeq
All of these gates are Gaussian, which means that the quantum algorithm can be implemented using Gaussian boson sampling devices. In the main text, we use the shorthand $\hat{D}(\bm{\alpha}), \hat{S}(\bm{r})$ with $\bm{\alpha}=(\alpha_1,\ldots,\alpha_m)$, $\bm{r}=(r_1,\ldots,r_m)$ to respectively denote displacement and squeezing gates acting individually on each mode, with $\alpha_i, r_i$ the displacement and squeezing parameters in mode $i$. 

\section{Molecular parameters}

We apply the Franck-Condon approximation \cite{franck1926elementary, condon1926a, condon1928nuclear} and assume that the change in the electronic state leads to a vertical transition between the vibrational energy levels of the two different electronic states of the molecule. The relation between the vibrational normal coordinates of these different electronic states is described by the Duschinsky transformation \cite{duschinsky1937}
\beq
\bm{q'} = U_D\bm{q} + \bm{d},
\eeq
where $U_D$ is the Duschinsky matrix, which is related to the overlap between the normal modes, and $\bm{d}$ is a real vector that describes the change in the molecular geometries of the electronic states.
The Duschinsky matrix is obtained from the eigenvectors of the Hessian matrices of the initial and final electronic states, $\bm{L}$ and $\bm{L}'$, respectively \cite{ezSpectrum, jahangiri2020quantum}
\beq\label{Eq: UD}
U_D = (\bm{L}')^T \bm{L}.
\eeq
The displacement vector $\bm{d}$ is obtained from the the Cartesian geometry vectors of the initial and final electronic states, $\bm{x}$ and $\bm{x}'$, as \cite{ezSpectrum, jahangiri2020quantum}
\beq
\bm{d} = (\bm{L}')^T m^{1/2} (\bm{x} - \bm{x}'), 
\eeq 
where $m$ is a diagonal matrix of atomic masses. The quantities $\bm{L}$, $\bm{L}'$, and $\bm{x}$ were obtained from density functional theory (DFT) calculations performed on both electronic states of MgP. The Duschinsky matrix for MgP is plotted in Fig.~\eqref{fig:dus}. The off-diagonal non-zero terms in this matrix are due to the mixing of the vibrational modes of the initial and final electronic states.        

The Duschinsky matrix and displacement vector are used to program the quantum computer with parameters that determine the unitary operator $\hat{U}_{\text{Dok}}$. This operator is defined in terms of displacement $\hat{D}(\bm{\alpha})$, squeezing $\hat{S}(\bm{r})$, and rotation $\hat{R}(U_L)$, $\hat{R}(U_R)$ operations \cite{huh2015boson, quesada2019franck}
\beq\label{Eq: Dok_Gaussian}
\hat{U}_{\text{Dok}} = \hat{D}(\bm{\alpha})\hat{R}(U_L) \hat{S}(\bm{r}) \hat{R}(U_R),
\eeq
where $U_L$ and $U_R$ are unitary matrices. The matrices $U_L$, $U_R$, and the vector $\bm{r}$ are obtained from the singular value decomposition $J = U_L\,\text{diag}(\bm{r})\, U_R$ of the matrix $J:=\Omega' U_D\Omega^{-1}$ where \cite{huh2015boson}
\begin{align}\label{Eq: omega}
\Omega &= \text{diag} (\sqrt{\omega_1},...,\sqrt{\omega_M}),\\
\Omega' &= \text{diag} (\sqrt{\omega_1'},...,\sqrt{\omega_M'}).
\end{align}
The terms $\omega$ and $\omega'$ in Eq. \eqref{Eq: omega} are the vibrational frequencies of the two electronic states which are also obtained from DFT calculations. 
The displacement vector $\bm{\alpha}$ in Eq.~\eqref{Eq: Dok_Gaussian} is obtained from the Duschinsky displacement vector as $\bm{\beta}=\hbar^{-1/2}\Omega'\bm{d}/\sqrt{2}$
where $\hbar$ is the reduced Planck constant.

\section{Complexity}

The number of optical modes in the quantum computer is equal to the number of vibrational modes, so the algorithm has linear space complexity. The squeezing and displacement gates act individually on each mode, leading to a constant depth circuit. The largest set of operations occur in the interferometer, where it is known that arbitrary interferometers can be decomposed into circuits of linear depth~\cite{reck1994}. Therefore, the complexity of generating a single sample scales linearly with problem size.

Now consider the number of samples needed. The averages of the Fermi function
\beq
\bar{f}(i, V)=\int_{\epsilon_{i}-\Delta}^{\epsilon_i+\Delta} d\epsilon\;f(\epsilon, V),
\eeq
are one-dimensional integrals that can be computed efficiently. In fact, since they are independent of the molecule, they can be pre-computed and loaded from memory. To compute transfer rates, we need to estimate $M$ probabilities $q(i)$. Each such estimation can be viewed as a Bernoulli trial where success corresponds to observing a sample inside of the energy subinterval $[\epsilon_i-\Delta, \epsilon_i+\Delta]$. In the limit of a large number of samples $N$, the unbiased estimator $\hat{q}(i)=\frac{N_i}{N}$ follows a normal distribution with standard deviation $\sigma=\sqrt{\frac{q(i)[1-q(i)]}{N}}$, which can be used to quantify the error in the estimation as
\begin{align}
\varepsilon &:=\frac{\sigma}{q(i)}\approx \sqrt{\frac{1}{N q(i)}},
\end{align}
where the approximation holds when $q(i)$ is small. Since there are only $M$ probabilities $q(i)$, the average probability is exactly $1/M$. Probabilities that are much smaller than $1/M$ don't contribute significantly to the transfer rates. Therefore, in terms of the target error $\varepsilon$ and the algorithm parameter $M$, the number of samples needed scales as
\beq
N=O\left(\frac{M}{\varepsilon^2}\right),
\eeq
meaning polynomially many samples are sufficient. 

\bibliographystyle{apsrev}
\bibliography{ET_references}